\documentclass[a4paper,11pt]{article}
\usepackage{a4wide}                     
\usepackage[english]{babel}
\usepackage{amsmath}                    
\usepackage{amssymb}                    
\usepackage{amsthm}
\usepackage{bbold}
\usepackage[dvips]{color}
\usepackage{url}                        
\usepackage{graphicx}                   
\usepackage{subfigure}
\usepackage[small,bf,hang]{caption2}    
\usepackage{xspace}                     
\usepackage[latin1]{inputenc}           
\usepackage{float}                      
\usepackage{flafter}                    
\usepackage{listings}                   
\usepackage{marvosym}                   
\usepackage{textcomp}                   
\usepackage{fancyhdr}                   
\usepackage[nottoc]{tocbibind} 					
\usepackage{longtable}
\usepackage{pdfpages}  									
\usepackage[top=3cm, bottom=3cm, left=2cm, right=2cm]{geometry}
\usepackage{framed}
\usepackage{verbatim}
\usepackage{setspace}
\usepackage{color}

\usepackage{braket}
\usepackage{upgreek}
\usepackage{authblk}

\begin{document}


\title{Integration of Single Photon Emitters in 2D Layered Materials with a Silicon Nitride Photonic Chip}


\author[1,2]{Fr\'ed\'eric Peyskens}
\author[1]{Chitraleema Chakraborty}
\author[2]{Muhammad Muneeb}
\author[2]{Dries Van Thourhout}
\author[1]{Dirk Englund}
\affil[1]{Quantum Photonics Group, Research Laboratory of Electronics, Massachusetts Institute of Technology, Cambridge, Massachusetts 02139, USA}
\affil[2]{Photonics Research Group, INTEC, Ghent University-imec, Center for Nano- and BioPhotonics, Ghent University, Technologiepark-Zwijnaarde 126, 9052 Ghent, Belgium}

\maketitle

\section{Abstract}

Photonic integrated circuits (PICs) enable miniaturization of optical quantum circuits because several optic and electronic functionalities can be added on the same chip. Single photon emitters (SPEs) are central building blocks for such quantum circuits and several approaches have been developed to interface PICs with a host material containing SPEs. SPEs embedded in 2D transition metal dichalcogenides have unique properties that make them particularly appealing as PIC-integrated SPEs. They can be easily interfaced with PICs and stacked together to create complex heterostructures. Since the emitters are embedded in a monolayer there is no total internal reflection, enabling very high light extraction efficiencies without the need of any additional processing to allow efficient single photon transfer between the host and the underlying PIC. Arrays of 2D-based SPEs can moreover be fabricated deterministically through STEM patterning or strain engineering. Finally, 2D materials grown with high wafer-scale uniformity are becoming more readily available, such that they can be matched at the wafer level with underlying PICs. Here we report on the integration of a WSe$_{2}$ monolayer onto a Silicon Nitride (SiN) chip. We demonstrate the coupling of SPEs with the guided mode of a SiN waveguide and study how the on-chip single photon extraction can be maximized by interfacing the 2D-SPE with an integrated dielectric cavity. Our approach allows the use of optimized PIC platforms without the need for additional processing in the host material. In combination with improved wafer-scale CVD growth of 2D materials, this approach provides a promising route towards scalable quantum photonic chips.

\section{Introduction}

Photonic integrated circuits (PICs) enable the miniaturizing of complex quantum optical circuits with large numbers of photonic devices connected with optimized insertion losses and phase stability. \cite{ref1} Photons in a PIC are routed in a single spatial mode of a low-loss single mode waveguide, consisting of a high index core surrounded by lower index cladding materials to provide confinement of the optical mode. Spatial mode matching, which is crucial for classical and quantum interference, can be nearly perfect for such an architecture. \cite{ref1} The use of PICs moreover allows integration of several functionalities on a single chip, including photonic cavities to enhance light-matter interaction, filters to block or select specific wavelengths, integrated photodetectors, etc. \cite{ref2} A central building block for such quantum photonic circuits are single photon emitters (SPEs). \cite{ref2} Over the past decade a variety of material systems have been investigated to create on-chip SPEs, including III-V quantum dots\cite{ref15}, carbon nanotubes\cite{ref16} and crystal colour centers such as the NV\cite{ref17} or SiV\cite{ref18} centers in diamond. 
\par
More recently, SPEs were discovered in monolayer transition metal dichalcogenides (TMDCs)\cite{ref3,ref4,ref5,ref6,ref7} and monolayer and multilayer hexagonal boron nitride (hBN) \cite{ref8,ref9}. It has been shown that nanoscale strain engineering can be used to scale up the creation of such 2D-SPEs \cite{ref20,ref21,ref22}, but integration with a PIC has not been achieved so far. This would however alleviate some important issues met with other approaches for quantum photonic applications. First of all, techniques to transfer 2D materials or stack them by Van der Waals epitaxy to create complex heterostructures are by now getting well established, enabling easy interfacing with high quality PICs. \cite{ref10,ref11,ref23} Secondly, it is possible to achieve very high light extraction efficiencies because the emitters are embedded in a monolayer, avoiding total internal reflection. This is a major issue for diamond and III-V based quantum technologies, where a separate photonic structure is typically made in the host material to allow efficient single photon transfer between the host and underlying PIC. This adds serious challenges because separate PICs have to be fabricated in the host material and moreover may require precise pick-and-place techniques to integrate both PICs together. \cite{ref17,ref31} Furthermore, 2D materials can easily be integrated with electrical contacts \cite{ref12} to ultimately enable all-electrical single photon generation over a broad spectrum \cite{ref13} or to tune the single photon wavelength and symmetry by the quantum-confined Stark effect \cite{ref19,ref39}. Finally, 2D materials grown with high wafer-scale uniformity are becoming widely available \cite{ref24,ref37,ref38}, such that they can be matched at the wafer level with underlying photonic circuitry.
\par
Here we study the integration of a WSe$_{2}$ monolayer onto a Silicon Nitride (SiN) chip and demonstrate the coupling of 2D-based single photon sources with the guided mode of a SiN waveguide. SiN PICs are a useful platform for routing photons that carry quantum information since they provide low-loss transmission in the visible and are available in a CMOS-fab. \cite{ref25} We discuss how integrated cavity-emitter systems, evanescently coupled to a waveguide, should be designed to optimize single photon extraction into the waveguide. As such the full potential of a high quality and CMOS-compatible PIC platform can be exploited without the need for stringent processing in the host material itself. In combination with wafer-scale growth of 2D materials, this provides a promising route towards scaling of quantum photonic circuits.

\begin{figure*}[htbp]
\centering
\includegraphics[width=\textwidth]{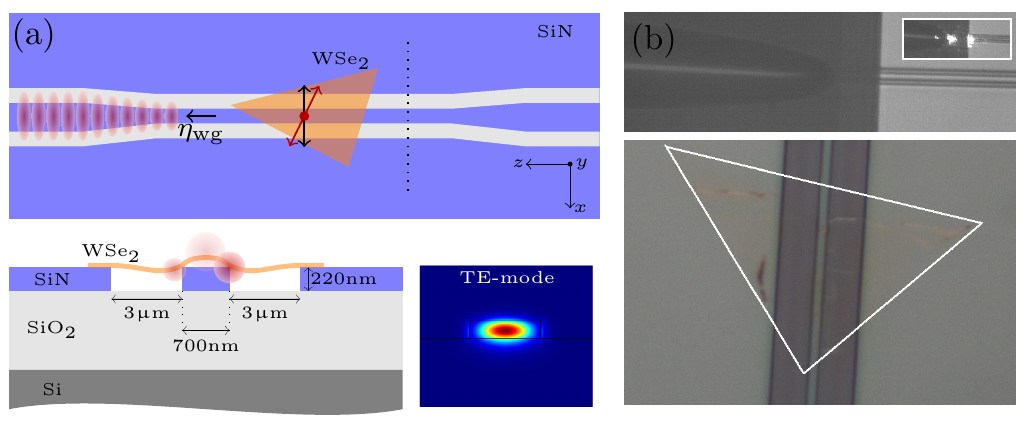}
\caption{\textbf{Integrated WSe$_{2}$ quantum emitters.} \textbf{(a)} Top view of the device: a WSe$_{2}$ flake is integrated on a 220 nm thick single mode SiN waveguide, separated by 2 air trenches from the bulk SiN. The waveguide ends are tapered to allow easier coupling with a lensed fiber. The orientation of the dipole moment of the WSe$_{2}$ emitters (red arrow) is random with respect to the quasi-TE polarization (approximately aligned along $x-$direction) of the fundamental waveguide mode (black arrow). A fraction $\eta_{\text{wg}}$ of the total emission couples into the left-propagating waveguide mode (represented by red shaded areas). Bottow left: Cross-section of the sample. The width of the air trenches and waveguide is 3 $\upmu$m and 700 nm respectively. The generated PL of emitters near the waveguide couples both to free-space and to the waveguide (red shaded circles). Bottom right: Cross-sectional mode profile (at $\lambda=750$ nm) of the waveguide, taken along the dotted black line in the top figure. \textbf{(b)} Top: Impression of the fiber-coupled chip (inset shows light coupling from the fiber to the chip). Bottom: microscope image of SiN chip with WSe$_{2}$ transferred on waveguide region. The flake is highlighted by the white triangle. 
}
\label{Fig1}
\end{figure*}

\section{Results}
\subsection{Device overview}

Figure \ref{Fig1}(a) shows a schematic of the device. A mechanically exfoliated WSe$_{2}$ flake is transferred by dry transfer onto a single mode SiN waveguide. After transfer, the sample was placed in an optical cryostat from Montana instruments and cooled down to 3.9K. Photoluminescence (PL) from the WSe$_{2}$ can either couple to free-space radiation or to the guided mode of the waveguide. The radiation to free-space is collected by a top objective with NA=0.65, while the waveguide-coupled PL is captured by a lensed fiber, aligned to the output facet of the waveguide. An impression of the fiber-coupled chip and a microscope image of the integrated WSe$_{2}$ flake are depicted in Figure \ref{Fig1}(b). Figure \ref{Fig1}(c) shows a SEM image of the SiN waveguide. See Supplementary Information for more information on the device fabrication and experimental setup. 
\par
To maximize the count rate of an integrated single photon source, the fraction $\eta_{\text{wg}}$ of total PL that couples to the waveguide mode should be as close as possible to one. It is however impossible to achieve this with the simple waveguide geometry shown in Figure \ref{Fig1}(a), but interaction with a cavity can significantly boost the overall coupling rate to the guided mode. As an extension of our experimental results we will therefore investigate for which cavity parameters near-unity waveguide extraction efficiencies can be obtained. An essential parameter in this calculation is the cavity-emitter coupling, which critically depends on the dipole moment strength of the integrated 2D-based emitter. For realistic estimates of this value, we will assess it from our experiments. As such we can get a clear overview of which cavity $Q-$factors and mode volumes $V_{c}$ are required to maximize single photon extraction.

\subsection{On-chip quantum emitters}

\begin{figure*}[htbp]
\centering
\includegraphics[width=\textwidth]{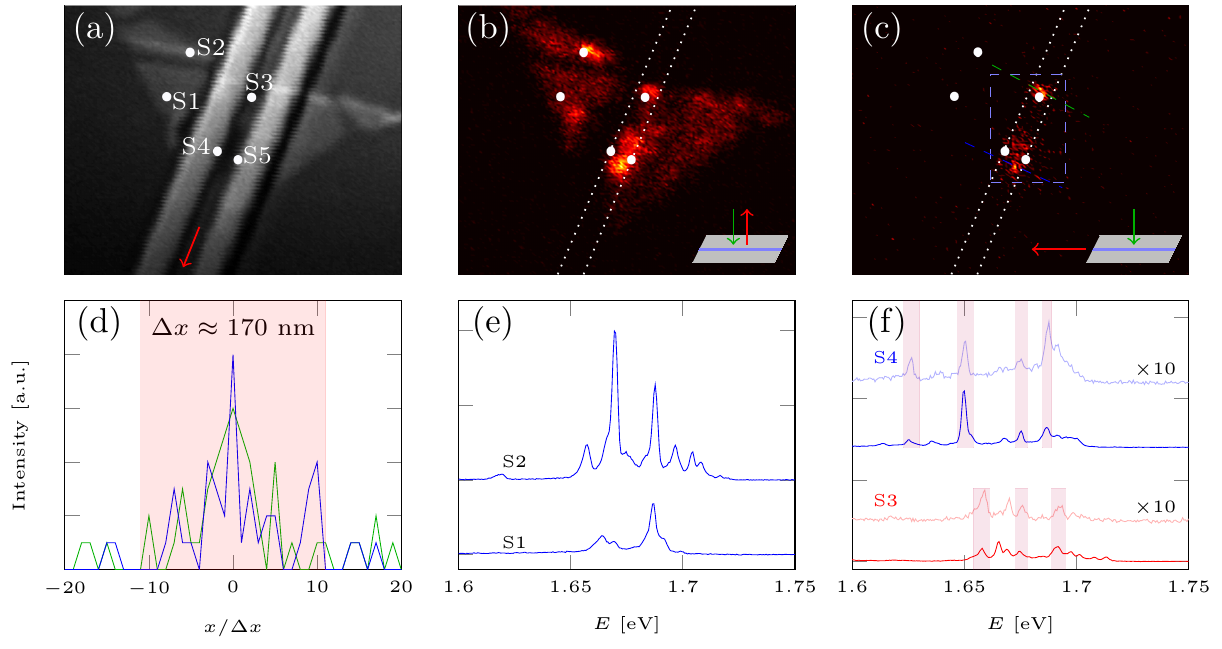}
\caption{\textbf{Waveguide-coupled WSe$_{2}$ quantum emitters.} \textbf{(a)} Confocal laser scan ($\lambda=532$ nm) of the relevant sample area. Spots S1 and S2 are spots off the waveguide, while spots S3 to S5 mark positions on the waveguide. The red arrow indicates the direction along which the fiber-coupled light is collected. \textbf{(b)} Confocal PL scan, by scanning the excitation beam over the sample from the top and collecting the PL from the top. \textbf{(c)} Waveguide PL scan, by scanning the excitation beam over the sample from the top and collecting the PL through the fiber. The white dotted lines mark the waveguide position. \textbf{(d)} Line scan along the the green and blue lines in Figure (c), highlighting the estimated spatial region coupled to the waveguide (shaded red region). \textbf{(e)} PL spectra from spots S1 and S2, collected from the top. \textbf{(f)} PL spectra from spots S3 (red) and S4 (blue), collected from the top (solid color) and through the fiber (shaded color). Matching peaks are highlighted by shaded purple regions. Where necessary, the spectrum baseline is shifted for improved visualization. The waveguide-coupled spectra are multiplied by 10. The excitation power for all PL spectra was $P_{e}\approx 25$ nW.}
\label{Fig2}
\end{figure*}

Figure \ref{Fig2} summarizes PL measurements on the flake. The excitation beam  ($\lambda=532$ nm) can be scanned over the sample through the top window of the cryostat by a set of two galvo-mirrors. The regions that light up in the PL scan of Figure 2(b), match with the area covered by the flake in the scanning confocal image of Figure 2(a). We will investigate five different spots on the flake, labeled S1 through S5. The spectra for 2 positions off the waveguide (S1 and S2) are shown in Figure \ref{Fig2} (d). Spot S1 exhibits only two prominent peaks, which are relatively weaker compared to the spot S2 peaks. Spot S2, which appears to be near the monolayer edge as evidenced by both the confocal and PL scan, contains several narrower peaks with FWHM on the order of 3 meV in the 1.65 to 1.7 eV spectral region. This result is similar to observations made by Tonndorf et al.\cite{ref3}. For all spectra in Figure \ref{Fig2}, the excitation power was set to 25 nW with an integration time of 60 seconds. Because the excitation power was low, the FWHM was not affected by power broadening. Spectral wandering during the long integration time could however result in inhomogeneous broadening of the FWHM of the emitters, as observed in earlier studies. \cite{ref7}
\par
The areas near spots S3, S4 and S5 exhibit brighter fluorescence compared to the surrounding region (see Figure \ref{Fig2} (b)) and are all located in the vicinity of the waveguide (region between the white dotted lines). This is similar to recent reports in which bright emission of a TMDC monolayer was observed at positions where the material was bend over a nanopillar and hints to the presence of strain-induced emitters coupled to the waveguide. \cite{ref21,ref22} To confirm that these spots are indeed coupled to the waveguide, we scan the excitation beam from the top, but collect the PL through the lensed fiber and indeed observe that only the waveguide region lights up (Figure \ref{Fig2}(c)). A small offset in the piezo position of the fiber from the waveguide results in an immediate loss of the signal, further confirming that we indeed collect light originating from the waveguide. Figure \ref{Fig2}(d) shows a line scan along two lines perpendicular to the waveguide to estimate the spatial extent over which the PL can couple into the waveguide. Emitters located up to $1.9 \upmu$m on either side of the waveguide can couple into the waveguide. A closer examination of the confocal and waveguide-coupled spectra of spots S3 and S4 is shown in Figure \ref{Fig2} (e). The spectra feature several narrow lines, with a typical linewidth ranging between 2.5 meV and 4 meV. This linewidth can be significantly broadened by the immediate surrounding of the WSe$_{2}$ (e.g. surface charges in the SiO$_{2}$ and SiN), but the broadening can be partially alleviated by encapsulation with hBN. \cite{ref26,ref40} A comparison between the spectrum of spot S1 and the other spots moreover shows more peaks near the waveguide or cracks in the sample, substantiating the argument that the emitters are indeed strain-induced. Data from a hyperspectral scan of the blue-dashed area in Figure \ref{Fig2}(c), containing info on the spectral distribution of the PL and an estimation on the number of peaks, are included in the Supplementary Information.
\par
A comparison of the confocal and waveguide-coupled spectra shows that not all peaks appearing in the confocal spectra are present in the waveguide-coupled spectra. This can be understood from the fact that the coupling between the waveguide mode $\textbf{E}_{wg}$ (quasi-TE-mode in our case) and the dipole moment of the quantum emitter $\textbf{p}_{d}$ scales according to $\textbf{E}_{wg}\cdot\textbf{p}_{d}\propto\cos\theta_{d}$, with $\theta_{d}$ the angle between $\textbf{E}_{wg}$ (black arrow in Figure \ref{Fig1}(a)) and $\textbf{p}_{d}$ (red arrow in Figure \ref{Fig1}(a)). Hence, when $\theta_{d}\rightarrow \frac{\pi}{2}$, the coupling vanishes. According to numerical simulations with Lumerical FDTD solutions, about $\eta_{\text{wg}}=7.3\%$ of the total power radiated by a dipole (at $E=1.63$ eV) with $\theta_{d}=0$ and centered on the top surface of the waveguide couples in the left-propagating guided TE-mode. For the same dipole emitter, $\eta_{\text{NA}}\approx 6.5\%$ radiates upwards in an NA=0.65. A dipole at the same position on the waveguide but with $\theta_{d}=\frac{\pi}{2}$ doesn't radiate into the TE-mode (as expected by the $\cos\theta_{d}$ behaviour), while emitting $\approx 7.3\%$ upwards in an NA=0.65. So regardless of the orientation of the dipole, we expect about $7\%$ of the total radiation to be captured in an NA of 0.65, while the light captured by the waveguide heavily depends on $\theta_{d}$. As such, the large spread in relative strength between the confocal and waveguide-coupled signal of a certain peak stems from the fact that their ratio scales as $\eta_{\text{wg}}/\eta_{\text{NA}}\propto\cos\theta_{d}$. The relative strength between different peaks depends both on the dipole polarization as well as on the absolute dipole moment of the emitter.


\subsection{Waveguide-coupled single photon source}
\begin{figure*}[htbp]
\centering
\includegraphics[width=\textwidth]{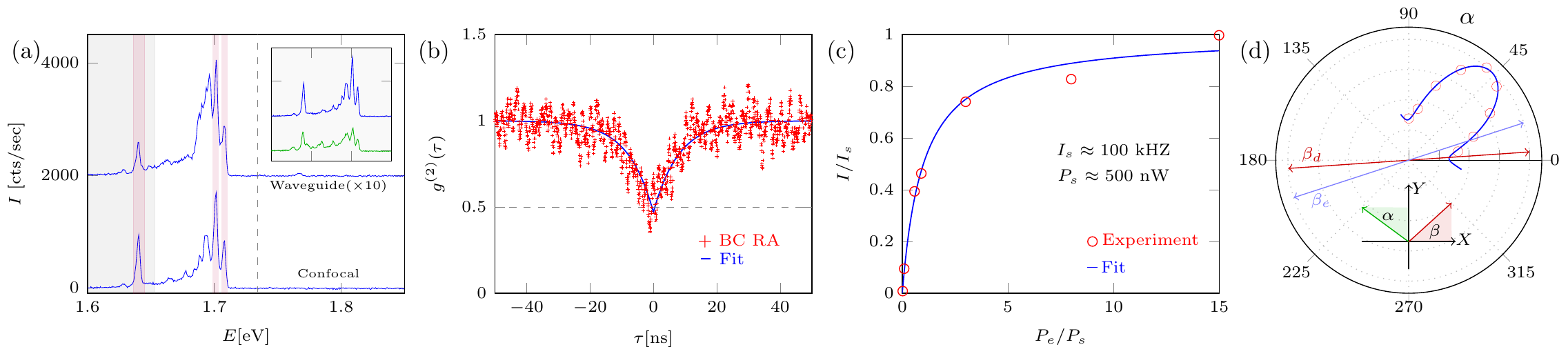}
\caption{\textbf{On-chip single photon emission.} \textbf{(a)} Confocal and waveguide-coupled spectrum of spot S5, excited with $\lambda=702$ nm. The waveguide spectrum is multiplied by 10 and offset by 2000 cts/sec for improved visualization. Common peaks are highlighted by the shaded purple regions. A 715 nm (1.73 eV) longpass filter, marked by the dashed line, was used to filter the pump. For the $g^{(2)}(\tau)$ measurement a 750 nm (1.65 eV) longpass filter (gray shaded area) was used to isolate the single emitter at 756.6 nm (1.64 eV). The inset figure shows confocal spectra obtained by either green ($\lambda=532$ nm) excitation (green curve) or excitation with $\lambda=702$ nm (blue curve). \textbf{(b-d)} Characterization of the 1.64 eV emitter. \textbf{(b)} Normalized background-corrected (BC) running average (RA) coincidence counts (red) and $g^{(2)}(\tau)$ fit (blue). \textbf{(c)} Measured intensity saturation (red) and fit to saturation curve (blue).\textbf{(d)} Normalized SPD count of the emitter (red) as a function of half-wave plate rotation angle $\alpha$ and fit to intensity transmission curve (blue); $\alpha=0$ corresponds to a half-wave plate fast axis along the $Y-$direction. See Supplementary Information for orientation of the half-wave plate with respect to the $(X,Y,Z)$ axes. Based on the fit, the difference in polarization angle between the PL ($\beta_{d}$) and excitation ($\beta_{e}$) beam can be extracted; $\beta=0$ corresponds to a polarization along the $X-$axis.}
\label{Fig3}
\end{figure*}
We will now focus on spot S5 of Figure \ref{Fig2}(a) and investigate the quantum nature of the observed emitters in more detail. The confocal and waveguide-coupled spectrum of spot S5 are shown in Figure \ref{Fig3}(a). We observe a few peaks recurring in both the confocal and waveguide spectrum, confirming that these emitters are indeed coupled to the waveguide. A prominent and isolated waveguide-coupled peak (FWHM$\approx 3$ meV) appears around 1.64 eV (756.5 nm). It has been shown that the PL of 2D-based quantum dots can be enhanced when the excitation laser wavelength is tuned close to the free excitonic resonance. \cite{ref4} When we scan the excitation wavelength with a tunable Ti:saph laser around the free exciton wavelength, we also find a considerable increase in peak count rate and reduction in background compared to excitation with $\lambda=532$ nm for the same excitation power (see inset Figure \ref{Fig3}(a)). An excitation wavelength of $\lambda=702$ nm provided the most optimal ratio between peak count rate and background, and hence the emitter was pumped at this wavelength for all subsequent experiments.
\par
A 750 nm longpass filter (gray shaded area in Figure \ref{Fig3}(a)) was used to spectrally isolate the 1.64 eV peak from the broad PL emission around 1.7 eV before the beam hits the Single Photon Detectors (SPDs). As such, the major contribution to the SPD count stems from the 1.64 eV peak and we can perform a $g^{(2)}-$measurement to investigate whether single photons are emitted by this emitter. Due to the lower count rates of the waveguide-coupled PL, we use the free-space collected PL for the $g^{(2)}-$measurement. Based on the spectrum we assess that the peak of interest (at 1.64 eV) contributes a fraction of about $\rho=0.76$ to the total signal while the rest is due to uncorrelated background. The raw normalized coincidence counts without any background correction are reported in the Supplementary Information, while the plot in Figure \ref{Fig3}(b) shows the background-corrected $g^{(2)}(\tau)-$curve, on which moreover a running average is applied to reduce the noise on the data. The background corrected $g^{(2)}_{BC}(\tau)$ value can be calculated according to $g^{(2)}_{BC}(\tau)=(g^{(2)}(\tau)-(1-\rho^{2}))/\rho^{2}$.\cite{ref27} See Supplementary Information for more details on the background correction and running average. Fitting the background-corrected data to the equation $g^{(2)}_{f}(\tau)=1-A\exp\left(-|\tau|/\tau_{f}\right)$ yields $g^{(2)}_{f}(0)=1-A=0.47$ and $\tau_{f}=7.99$ ns. \cite{ref41} The minimum value in the background-corrected data without averaging is about 0.03, which would hint to almost perfect single photon emission. The fitted rise time $\tau_{f}=7.99$ ns is a lower limit for the PL decay time and is in the same order of magnitude as previously reported values for WSe$_{2}$. \cite{ref3} The clear anti-bunching dip with a background corrected $g^{(2)}(0)<0.5$ confirms that the emitter indeed emits single photons. 
\par
A generic two-level system moreover exhibits saturation of the PL emission when the excitation rate increases, and this has been observed for WSe$_{2}$ emitters before. \cite{ref3,ref4,ref5,ref7} The PL saturation for our waveguide-coupled quantum emitter is shown in Figure \ref{Fig3}(b). A fit of the PL intensity $I=I_{s}\left(P_{e}/(P_{e}+P_{s})\right)$ as a function of excitation power $P_{e}$ yields a saturation power of $P_{s}\approx 500$ nW (at $\lambda=702$ nm) and a saturation intensity of $I_{s}\approx 100$ kHz. The excitation efficiency of the emitter will however depend on the orientation between the dipole moment of the quantum emitter $\beta_{d}$ and the excitation polarization $\beta_{e}$ and will hence affect the measured intensity. We therefore perform polarization-dependent transmission measurements to determine $\Delta\beta=\beta_{d}-\beta_{e}$. The normalized transmitted emitter count rate to SPD1 as a function of the polarization-rotating half-wave plate angle $\alpha$ is shown in Figure \ref{Fig3}(d). By fitting this count rate one can determine $\Delta\beta$ and eventually assess the saturation count rate of the single photon source. When corrected for transmission and collection efficiencies of the system, the total saturation intensity is about 3 MHz (to all modes, guided and non-guided) while the estimated maximum waveguide-coupled count rate is about 100 kHz (see Supplementary Information). Further improvements consist of changes in the waveguide design \cite{ref28} or interaction with plasmonic or dielectric cavities \cite{ref29,ref30} to maximize the coupling efficiency into the guided mode and enhance non-classical light generation. 

\section{Optimizing on-chip single photon extraction and indistinguishability}

\begin{figure*}[htbp]
\centering
\includegraphics[width=\textwidth]{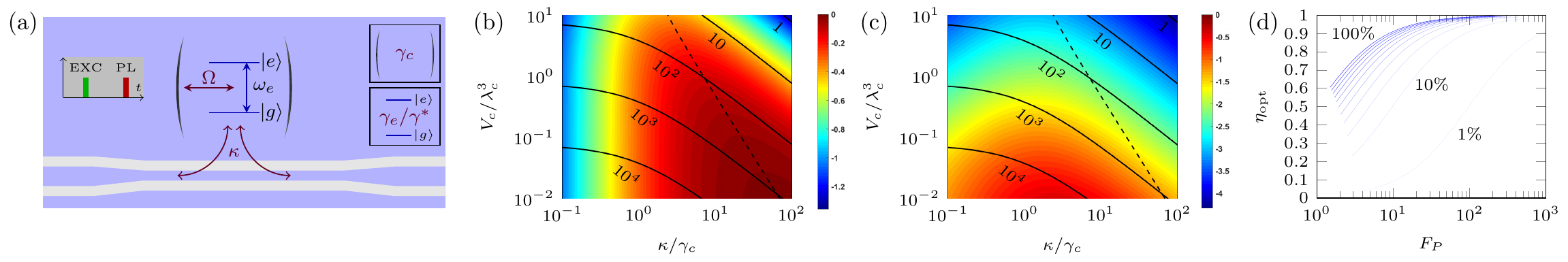}
\caption{\textbf{Integrated cavity-emitter system} \textbf{(a)} Schematic of an integrated cavity-emitter system, evanescently coupled to a single mode waveguide. The coupling rate between an emitter with frequency $\omega_{e}$ and a cavity with resonance frequency $\omega_{c}$ is given by $\Omega$. The decay rate from the cavity-emitter system to the guide mode is $\kappa$, while the other decay channels of the emitter and cavity are given by $\gamma_{e}$and $\gamma_{c}$ respectively. The emitter dephasing is described by $\gamma^{*}$. The system is excited (EXC) by a short pulse and subsequently the single photon PL is collected. \textbf{(b-c)} Single photon extraction efficiency $\eta$ \textbf{(b)} and extraction-indistinguishability product $\eta V$ \textbf{(c)} as a function of cavity mode volume $V_{c}$ and cavity decay rate $\kappa$. The black solid lines represent lines of constant Purcell factor $F_{P}$, while the black dashed line represents ($V_{c}$,$\kappa$) combinations for which $\eta$ is maximal. The parameter values used to generate plots \textbf{(b-c)} are $Q_{i}=10000$, $\Gamma=3$ MHz, $\gamma_{e}=100$ MHz, $\cos\theta_{d}=1$ (i.e. perfect alignment of the emitter polarization with the cavity field), $\gamma^{*}=100$ GHz \cite{ref35}, $n_{d}=4$ and $\lambda_{c}=\lambda_{0}/n_{\text{eff}}$ with a free-space wavelength of $\lambda_{0}=750$ nm and an effective refractive index of $n_{\text{eff}}=1.6$ for the fundamental TE-mode. Both plots are on a $\log_{10}$ color scale, i.e. 0 corresponds to perfect $\eta=1$ or $\eta V=1$. \textbf{(d)} Optimum $\eta_{\text{opt}}$ (evaluated at ($V_{c}$,$\kappa$) combinations for which $\eta$ is maximal) as a function of $F_{P}$ for different values of $\Gamma$ ranging from $0.01\gamma_{e}$ to $\gamma_{e}$ (different quantum yields).}
\label{Fig4}
\end{figure*}

Apart from high single-photon extraction efficiency, various applications (linear optical quantum computing, quantum teleportation, quantum networks, etc.) require the single photons to be indistinguishable (i.e. identical spatial and spectral modes). \cite{ref32} For an ideal single photon source, the product of extraction efficiency $\eta$ and indistinguishability $V$ should be $\eta V=1$. In this section we will assess how $\eta$ and $\eta V$ of an integrated 2D quantum emitter can be optimized by cavity coupling. Figure \ref{Fig4}(a) shows a schematic of the investigated platform. The emitter is coupled to a cavity with coupling strength $\Omega$, while the cavity is evanescently coupled to the waveguide with a coupling strength $\kappa$. The decay rate $\gamma_{c}$ represents intrinsic absorption losses and radiation to non-guided modes, while the rate $\gamma_{e}$ incorporates decay of the emitter to all modes (radiative and non-radiative) other than the cavity and $\gamma^{*}$ is the emitter dephasing. For our calculations we assume the emitter is resonant with the cavity ($\omega_{e}=\omega_{c}$) and is initialized in the excited state by a short excitation pulse (EXC) with no photons present in the cavity. The master equation governing the dynamics of this system is discussed in the Supplementary Information. In the regime where $\gamma^{*}\ll\gamma_{e}+\gamma_{p}$ (which should be satisfied for low temperatures and moderate $Q-$factor cavities), the single photon extraction efficiency into the guided mode ($\eta$) is given by
\begin{equation}
\eta=\frac{\kappa}{(\gamma_{e}+\gamma_{c}+\kappa)\left(1+\frac{\gamma_{e}(\gamma_{c}+\kappa)}{4\Omega^{2}}\right)}.
\end{equation}
The expressions for the indistinguishability $V$ of photons coupled into the guided mode, as derived by Grange et al.\cite{ref32}, depend on the regime within which the system falls (see Supplementary Information). To assess $\eta$ and $\eta V$ (as shown in Figure \ref{Fig4}(b-c)), we first need to determine the different coupling strengths.
The coupling constant $\Omega$ depends on the cavity mode volume $V_{c}$ through $\Omega^{2}=\frac{3\pi c^{3}}{2n_{d}\omega_{c}^{2}}\cos^{2}\theta_{d}\left(\frac{\Gamma}{V_{c}}\right)$, with $\Gamma$ the free-space radiative decay rate in a uniform dielectric with index $n_{d}$, and $\theta_{d}$ the angle between the emitter dipole moment and the cavity field. For our calculations we assume $n_{d}$ is the refractive index of a WSe$_{2}$ monolayer ($n_{d}=4$). \cite{ref33} In our case, the radiative decay rate to non-guided modes will usually differ from $\Gamma$ due to the non-uniform dielectric environment and may furthermore be influenced by the vicinity of the dielectric cavity, but as a simplifying assumption we set $\Gamma\approx\gamma_{r}$ with $\gamma_{r}$ the radiative decay rate determined from our experiment, i.e. $\gamma_{r}\approx 3$ MHz. We moreover assume perfect alignment between the emitter and cavity mode ($\cos\theta_{d}=1$). The decay rate $\gamma_{e}$ also contains contributions to non-radiative modes ($\gamma_{e}=\gamma_{r}+\gamma_{nr}$), and can be approximated by $\gamma_{e}=\gamma_{r}/\xi$ with $\xi$ the quantum yield of the monolayer. For exfoliated WSe$_{2}$, a quantum yield of $\xi\approx 3\%$ has been reported \cite{ref34}, such that $\gamma_{e}\approx 100$ MHz. The final parameter is $\kappa$, which we express through the intrinsic cavity quality factor $Q_{i}$ as $\kappa=\chi\gamma_{c}=\chi\left(\frac{\omega_{c}}{2Q_{i}}\right)=\frac{\omega_{c}}{2Q_{\kappa}}$ such that the loaded quality factor of the cavity is given by $Q=\left(Q_{i}^{-1}+Q_{\kappa}^{-1}\right)^{-1}=Q_{i}/(1+\chi)$. We use $Q_{i}=10000$ for our calculations. The above parameter values are now used to estimate how $\eta$ and $\eta V$ can be improved through cavity-assisted interaction as a function of the normalized cavity mode volume $V_{c}/\lambda_{c}^{2}$ and waveguide-cavity coupling $\chi=\kappa/\gamma_{c}$ (Figure \ref{Fig4}(b-c)). The solid black lines represent lines of constant Purcell factor $F_{P}=\frac{3}{4\pi^{2}}Q\left(\frac{\lambda_{c}^{3}}{V_{c}}\right)$, while the dashed black line represents the ($V_{c}$,$\kappa$) combinations for which $\eta$ is optimized. For a given mode volume $V_{c}$ (i.e. $\Omega$), the coupling rate $\kappa$ that maximizes $\eta$ is given by 
\begin{equation}
\kappa_{\text{opt}}=\gamma_{c}\sqrt{\left(1+\frac{\gamma_{e}}{\gamma_{c}}\right)\left(1+\frac{4\Omega^{2}}{\gamma_{e}\gamma_{c}}\right)}.
\end{equation}
For this value of $\kappa$, the optimum $\eta\approx \mathcal{C}/\left(1+\sqrt{1+\mathcal{C}}\right)^{2}$ if $\gamma_{e}<\gamma_{c}$, with $\mathcal{C}=4\Omega^{2}/(\gamma_{e}\gamma_{c})\propto \xi Q_{i}(\lambda_{c}^{3}/V_{c})$. As such, near-unity extraction requires a high intrinsic quality factor (while the loaded $Q$ can be much lower), high quantum efficiency and small mode volume. The intersection of the $F_{P}=100$ line with $\kappa_{\text{opt}}$ yields $\eta\approx 75\%$ for $\kappa=6.65\gamma_{c}$ ($Q\approx 1307$) and $V_{c}=\lambda_{c}^{3}$. For these parameter values, $\eta V$ is only $0.5\%$ however. To achieve high $\eta V$ one typically needs much smaller $V_{c}$ because the cooperativity $\mathcal{C}$ has to overcome the emitter dephasing $\gamma^{*}$. If we decrease $V_{c}$ to $V_{c}=0.01\lambda_{c}^{3}$, then a maximum $\eta V\approx 37\%$ is achieved for $\kappa=2.55\gamma_{c}$ ($Q\approx 2817$). A near-unity extraction ($\eta=95\%$) can be achieved for $\kappa=39.5\gamma_{c}$ ($Q\approx 247$) and $V_{c}=0.03\lambda_{c}^{3}$ ($F_{P}\approx 626$), with $\eta V\approx 3.4\%$. By using the ultrasmall mode volume nanocavities reported in \cite{ref36}, we could hence achieve near perfect single photon extraction, even for a very low quantum yield emitter. Nevertheless, the corresponding $\eta V$ product is still more than an order of magnitude away from the ideal value. A higher quantum yield could partially alleviate this issue and moreover allows near-unity $\eta$ for moderate Purcell factors as shown in Figure \ref{Fig4}(d), which depicts $\eta_{\text{opt}}$ (i.e. $\eta$ evaluated at ($V_{c}$,$\kappa$) combinations for which $\eta$ is maximal) as a function of $F_{P}$ and $\xi$. For near-unity quantum yield, $\eta$ already reaches $91\%$ for a moderate Purcell factor of $F_{P}=10$, while $\eta=99\%$ for $F_{P}=100$. If we on the other hand fix $V_{c}=0.01\lambda_{c}^{3}$, then the maximal $\eta V$ increases to $\eta V=84\%$ ($\eta=93\%$) for $\kappa\approx 12.5\gamma_{c}$ ($F_{P}\approx 5600$) and $\xi=1$. The combination of a cavity with mode volume $V_{c}=0.01\lambda_{c}^{3}$ and $Q_{i}=10000$ with a near-unity quantum yield 2D-emitter could hence approach the limit of an ideal single photon source ($\eta V=1$). This analysis can be repeated for any dielectric cavity-emitter system that is evanescently coupled to the waveguide and as such can guide future design efforts to optimize single photon extraction and indistinguishability of photons coupled into the guided mode of the waveguide. 

\section{Conclusion}

In conclusion we have demonstrated that integration of a WSe$_{2}$ monolayer onto a SiN waveguide results in quantum emitters evanescently coupled to the waveguide. Second-order correlation measurements on a spectrally isolated quantum emitter confirm that single photons are emitted with a waveguide-coupled saturation count rate of 100 kHz. These results confirm previous claims that strain-induced quantum emitters could be coupled to photonic structures. \cite{ref21,ref22} A numerical analysis on the optimization of single photon extraction and indistinguishability using integrated dielectric cavity-emitter systems indicates that near-unity single photon extraction can be achieved, even for low quantum yield emitters. The presented approach for integration of strain-induced TMDC-based SPEs retains the favorable attributes of SiN PICs without the need for stringent processing in the quantum emitter host material itself. Recent progress in wafer-scale growth and patterning of identical 2D-material based devices \cite{ref24,ref37,ref38} provides a promising route in combination with our waveguide-coupled 2D-SPEs to scale up quantum photonic circuits.


\section{Acknowledgements}
We acknowledge Liesbet Van Landschoot and Steven Verstuyft for processing of the SiN chips, Hyowon Moon for building the confocal setup, and Noel Wan for help in making the custom vacuum fiber feedthrough and installing the fiber-coupling unit. F.P. acknowledges support from an FWO (Fonds voor Wetenschappelijk Onderzoek - Vlaanderen) postdoctoral fellowship. D.E. and F.P. acknowledge partial support from the NSF EFRI-ACQUIRE program ``Scalable Quantum Communications with Error-Corrected Semiconductor Qubits'' and the Army Research Laboratory Center for Distributed Quantum Information (CDQI).


\section{Supplementary Information}

\subsection{Fabrication}
The silicon nitride waveguides were patterned using standard e-beam lithography on a Raith Voyager system. A positive e-beam resist ARP-6200.09 was spincoated (3000 RPM, baked at 150$^{\circ}$C for 1 min) on a commercially grown slab wafer (220 nm SiN on SiO$_{2}$ on Si). Subsequently a protective coating (Electra 92) was spincoated (2000 RPM, baked at 90$^{\circ}$C for 2 min) on top of the resist. Air trenches (3 $\upmu$m wide) were defined in FBMS mode (Fixed Beam Moving Stage) to avoid stitching errors at the boundaries of the writing field. The Electra 92 coating can be removed by DI water. After development and etching, a 700 nm wide waveguide remains between the air trenches. The samples were thoroughly cleaned before dry transfer. WSe$_{2}$ flakes were mechanically exfoliated from a bulk crystal purchased from hqgraphene. The flakes were subsequently picked up by a GelPak stamp and transferred to the SiN surface by gently releasing the GelPak stamp from the substrate.

\subsection{Experimental setup}

\begin{figure*}[htbp]
\centering
\includegraphics[width=0.7\textwidth]{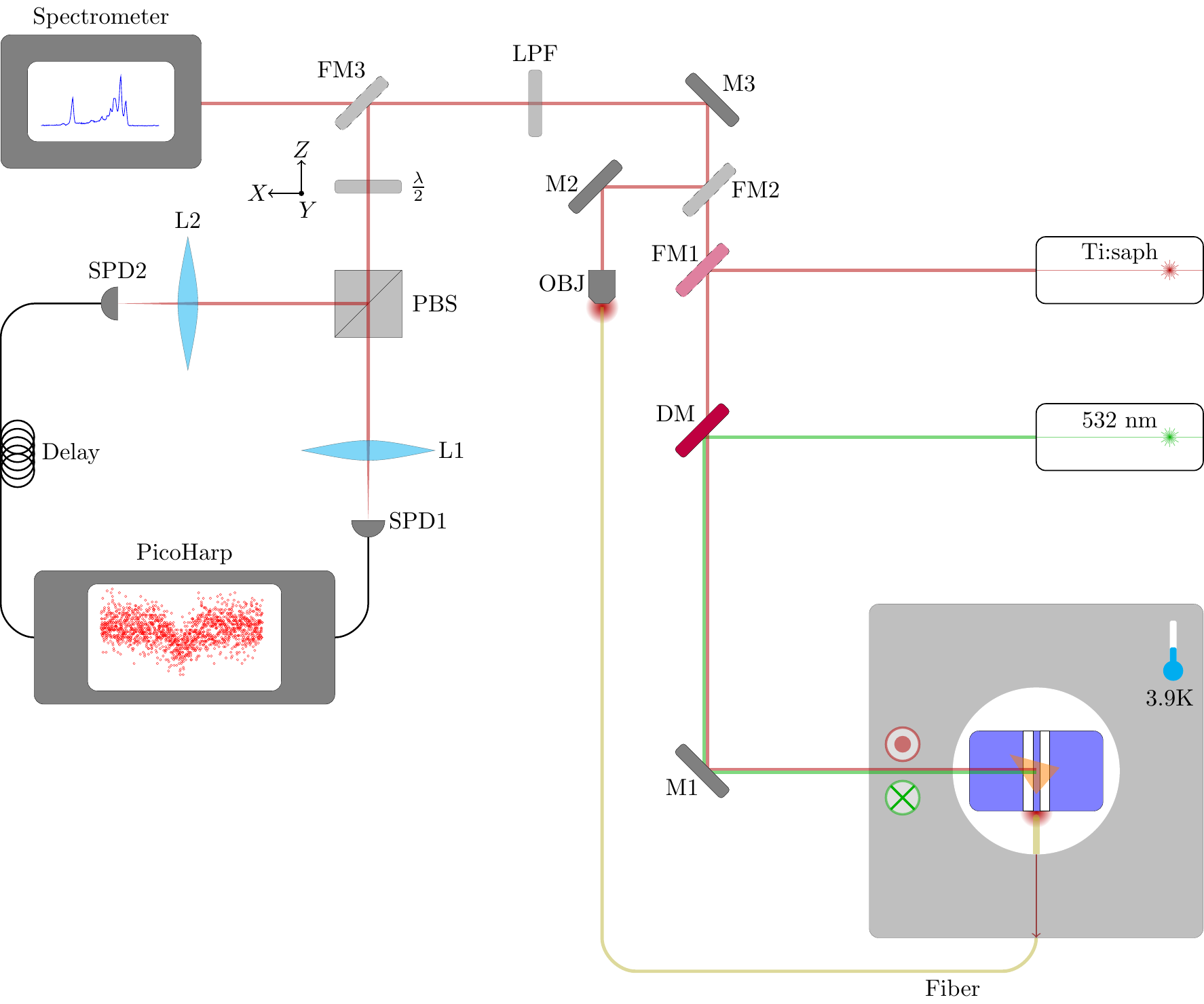}
\caption{\textbf{Measurement setup} M: fixed mirror, FM: flip mirror, DM: dichroic mirror, OBJ: fiber collimating objective, LPF: longpass filter (either for 532 nm or 715 nm), $\frac{\lambda}{2}$: half-wave plate, PBS: polarizing beamsplitter, L:focusing lens, SPD: Single Photon Detector} \label{FigS1}
\end{figure*}

The setup is shown in Figure \ref{FigS1}. The cryostat has a top window to excite the sample from above ($X$ symbol highlighting the propagation direction of the excitation beam) and through which the free-space PL is collected ($\circ$ symbol highlighting the propagation direction of the PL beam). A system of two galvo-mirrors allows to scan the excitation beam over the sample through the top window of the cryostat. One of the side ports is equipped with a home-built vacuum fiber feedthrough to minimize transmission losses between the cryostat and the outside world. Depending on the experiment, FM1 can be flipped to excite the sample with variable wavelengths from an Msquared Ti:saph laser. The dichroic mirror (DM) filters the green excitation beam from the PL. The PL collected in the fiber is sent to a fiber collimator unit and is subsequently coupled in the same path used for characterization of the free-space PL by flipping FM2 (this also hinders any free-space PL to be collected while we are studying the waveguide-coupled PL). A longpass filter (LPF) filters out remaining contributions from the pump beam. Without FM3, the PL is sent into a free-space spectrometer. When flipping FM3, the same beam is sent to an HBT setup for second-order correlation measurements. Combination of the half-wave plate with the polarizing beamsplitter allows to balance the counts on the 2 SPDs (the beams are focused by a lens on the surface of the free-space SPDs) and allows to study the relative polarization between the excitation and PL beam.

\subsection{Hyperspectral scan of the integrated emitters}

\begin{figure*}[htbp]
\centering
\includegraphics[width=0.7\textwidth]{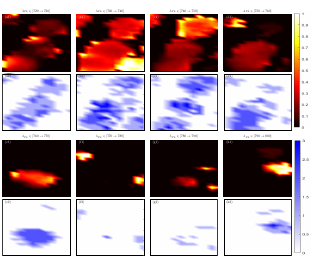}
\caption{\textbf{Hyperspectral scan.} \textbf{(a1-h1)} Normalized total spectral count and \textbf{(a2-h2)} number of peaks in the spectral band 720 to 800 nm. Each subplot contains info on a spectral subband of 10 nm wide.} \label{FigScan}
\end{figure*}

This sections contains data from a hyperspectral scan of an area near the waveguide (highlighted by the blue-dashed area in Figure 2(c) of the main text). For each point in the scan we took a spectrum, calculated the total spectral count and normalized the spectral count in different narrower subbands (each 10 nm wide) to this total count (Fig.\ref{FigScan}(a1-h1)). For each spatial point, we also assessed the number of clear peaks in the different spectral subbands (Fig.\ref{FigScan}(a2-h2)). One can see that the PL emission predominantly consists of peaks with a wavelength in the 720 nm to 760 nm range and that in each subband of 10 nm, the average number of peaks ranges from 1 to 3, i.e. an average of 4 to 12 peaks in the 720 nm to 760 nm wavelength region.

\subsection{Second order correlation measurements}

\begin{figure*}[htbp]
\centering
\includegraphics[width=\textwidth]{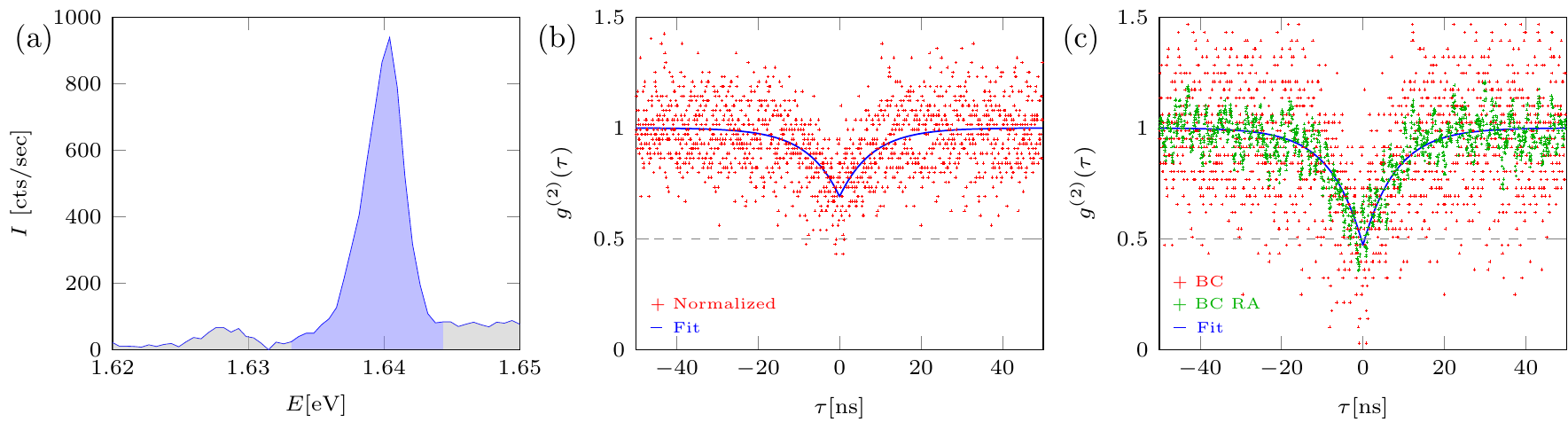}
\caption{\textbf{Second order correlation measurements.} \textbf{(a)} Assessment of the signal-to-background ratio $\rho$ (blue=signal, gray=background), based on the spectral count rate of emission with a wavelength above 750 nm. \textbf{(b)} Normalized coincidence counts (red) and fit to the normalized, uncorrected data. \textbf{(c)} Background-corrected BC (red) and background-corrected running average BC RA (green) coincidence counts with corresponding fit (blue).} \label{Figg2}
\end{figure*}

Figure \ref{Figg2}(a) shows the spectrum of spot S5 for wavelengths above 750 nm (below 1.65 eV). Based on this spectrum we assess that the peak of interest (at 1.64 eV) contributes a fraction of about $\rho=0.76$ (blue shaded area) to the total signal while the rest is due to uncorrelated background (gray shaded areas). The raw normalized coincidence counts are shown in Figure \ref{Figg2}(b), with a minimum value of 0.43. A fully unconstrained fit (in which we don't require that the minimum of the curve should equal 0.43) however yields $g^{(2)}(0)=0.69$ and $\tau_{f}=7.98$ ns. While the emission exhibits anti-bunching, the $g^{(2)}(0)$ value should drop below 0.5 as a clear sign of single photon emission. When applying background correction (BC) to the raw data, we obtain the red data shown in Figure \ref{Figg2}(c). For improved visualization, an $M$-point running average ($M=11$) was applied to reduce the noise (green data). These data are shown in the main text of the paper. The running average $g^{(2)}_{RA}(\tau)$ data at each time $\tau$ are obtained using the formula
\begin{equation}
	g^{(2)}_{RA}(\tau)=\frac{1}{M}\sum_{n=-\frac{(M-1)}{2}}^{\frac{M-1}{2}}g^{(2)}(\tau+nT),
\end{equation}
with $T$ the timing resolution of the measurement. For the background-corrected data, $g^{(2)}(0)=0.47$, confirming single photon emission.

\subsection{Brightness of the integrated single photon source}
The counts incident on SPD1 (which are used to assess the brightness), originate from a light field $\textbf{E}=E_{0}(\cos\beta\textbf{e}_{X}+\sin\beta\textbf{e}_{Y})$ that consecutively passes through a waveplate with orientation $\alpha$ and a polarizing beamsplitter PBS (for a definition of angles $\alpha$ and $\beta$, see Figure 3(d)). We first express this light field in the frame of the waveplate, which has basis vectors
\begin{align}
	\textbf{e}_{X'}&=\sin\alpha\textbf{e}_{X}-\cos\alpha\textbf{e}_{Y} \\
	\textbf{e}_{Y'}&=\cos\alpha\textbf{e}_{X}+\sin\alpha\textbf{e}_{Y},
\end{align}
such that
\begin{equation}
\textbf{E}=E_{0}(\sin(\alpha-\beta)\textbf{e}_{X'}+\cos(\alpha-\beta)\textbf{e}_{Y'}).
\end{equation}
In the frame of the wave-plate, the slow axis ($\textbf{e}_{Y'}$) obtains a $\pi$ phase shift, such that the field after the wave-plate and back in the original frame $(\textbf{e}_{X},\textbf{e}_{Y})$ is given by
\begin{equation}
\textbf{E}_{\text{aw}}=-E_{0}(\cos(2\alpha-\beta)\textbf{e}_{X}+\sin(2\alpha-\beta)\textbf{e}_{Y}).
\end{equation}
The PBS does not provide perfect filtering between the $\textbf{e}_{X}$ and $\textbf{e}_{Y}$ polarization, so we attribute a power transmission of $T_{X}$ and $T_{Y}$ to the respective components. So the intensity reaching SPD1 is eventually given by
\begin{equation}
I_{\text{SPD}}=I_{0}(T_{X}\cos^{2}(2\alpha-\beta)+T_{Y}\sin^{2}(2\alpha-\beta)),
\end{equation}
with $I_{0}$ the intensity of the original beam. A fit of the SPD1 signal counts as a function of $\alpha$ yields the following fitting values: $T_{X}=0.17$, $T_{Y}=0.51$ and $\beta_{d}=3.9^{\circ}$ for the dipole emitter and $\beta_{e}=18^{\circ}$ for the excitation polarization respectively. The deviation from the optimal excitation efficiency, i.e. $\beta_{d}=\beta_{e}$, is hence only $1-\cos(\beta_{d}-\beta_{e})\approx 3\%$. So both polarizations are well aligned for this particular emitter, and a negligible increase of $1/\cos(\beta_{d}-\beta_{e})\approx 1.03$ would be expected in the count rate if both polarizations were perfectly aligned.
\\
We moreover approximate the overall transmission loss due to all optics between the half-wave plate and the collection objective to be about $T_{o}=50\%$. As discussed in the main text, about $\eta_{NA}\approx 7\%$ of the total radiation is captured by the collection objective. The maximum SPD1 count rate $I_{\text{SPD}}^{\text{max}}\approx 29000$ cts/sec was obtained for $\alpha\approx 50^{\circ}$ and $P_{e}\approx P_{s}$. This implies $I_{0}\approx 50$ kHz and $I_{s}=2I_{0}\approx 100$ kHz. Taking into account the remaining transmission and collection losses, the estimated brightness of the single photon source is about $I_{0}/(T_{o}\eta_{NA})\approx 1.6$ MHz ($I_{s}\approx 3$ MHz). For an ideal dipole orientation and position, about $7\%$ couples into the forward propagating waveguide mode, which leads to an estimated maximum waveguide-coupled count rate of 100 kHz.

\subsection{Evanescently coupled cavity-emitter systems}

\subsubsection{Master equation}

We will describe the evanescently coupled dielectric cavity-emitter system by the same master equation as reported in our earlier work. \cite{Sref29,Sref30} In a frame rotating at the emitter frequency $\omega_{e}$, the density matrix $\rho$ satisfies
\begin{multline}
\frac{d\rho}{dt}= -i\Omega [pS_{+}+p^{\dagger}S_{-},\rho] + \frac{\gamma_{p}}{2}\left(2 p\rho p^{\dagger} - p^{\dagger}p\rho - \rho p^{\dagger}p \right) \\ +\frac{\gamma_{e}}{2}\left(2 S_{-}\rho S_{+} - S_{+}S_{-}\rho - \rho S_{+}S_{-} \right) \\ +\frac{\gamma^{*}}{2}\left(2 S_{z}\rho S_{z} - S_{z}S_{z}\rho - \rho S_{z}S_{z} \right),
\end{multline}
with the assumption that the cavity (described by the annihilation operator $p$) is resonant with the emitter (i.e. $\omega_{c}=\omega{e}$). The spin operators for the emitter satisfy $S_{+}=\ket{e}\bra{g}$, $S_{-}=\ket{g}\bra{e}$ and $S_{z}=\frac{1}{2}\left(\ket{e}\bra{e}-\ket{g}\bra{g}\right)$. The decay rates $\gamma_{p}$, $\gamma_{e}$ and $\gamma^{*}$ respectively represent the overall decay rate of the cavity (both due to intrinsic losses and decay into the waveguide), the decay rate of the emitter into the non-guided modes and the dephasing rate of the quantum emitter.
\par
The cavity-emitter coupling strength $\Omega$ is given by
\begin{equation}
	\Omega=\sqrt{\frac{\omega_{c}}{2\hbar\epsilon_{0}}}|\textbf{p}_{d}|\cos\theta_{d}\left(\frac{1}{\sqrt{V_{c}}}\right)
\end{equation}
with $V_{c}$ the cavity mode volume, $|\textbf{p}_{d}|$ the strength of the dipole moment of the emitter and $\theta_{d}=\arccos(\textbf{e}_{d}\cdot\textbf{e}_{c})$ the angle between the unit polarization vector of the emitter $\textbf{e}_{d}$ and the cavity field $\textbf{e}_{c}$. \cite{SrefB1} The mode volume is defined as
\begin{equation}
	V_{c}=\frac{\iiint d\textbf{r}\epsilon(\textbf{r})|\textbf{E}_{m}(\textbf{r})|^{2}}{\epsilon(\textbf{r}_{d})|\textbf{E}_{m}(\textbf{r}_{d})|^{2}},
\end{equation}
with $\textbf{E}_{m}(\textbf{r})$ the cavity mode field and $\epsilon(\textbf{r})$ the relative permittivity of the medium. The mode volume is normalized using the mode field at the position of the dipole emitter $\textbf{r}_{d}$ (and hence not using the maximum of the mode field). The strength of the dipole moment $|\textbf{p}_{d}|$ can be related to the emitter decay rate $\Gamma$ in a uniform dielectric with refractive index $n_{d}$ through
\begin{equation}
	\Gamma=\frac{n_{d}\omega_{e}^{3}|\textbf{p}_{d}|^{2}}{3\pi\epsilon_{0}\hbar c^{3}} \Rightarrow |\textbf{p}_{d}|=\sqrt{\frac{3\pi\epsilon_{o}\hbar c^{3}\Gamma}{n_{d}\omega_{e}^{3}}}
\end{equation}
such that
\begin{equation}
	\Omega^{2}=\frac{3\pi c^{3}}{2n_{d}\omega_{c}^{2}}\cos^{2}\theta_{d}\left(\frac{\Gamma}{V_{c}}\right).
\end{equation}
The decay rate $\gamma_{p}=\gamma_{c}+\kappa$ consists of the intrinsic decay rate of the cavity $\gamma_{c}=\frac{\omega_{c}}{2Q_{i}}$ (determined by the intrinsic quality factor $Q_{i}$ which includes absorption and radiation losses to non-guided modes) and the coupling rate to the guided modes $\kappa$. We assume that $\kappa=\chi\gamma_{c}=\frac{\omega_{c}}{2Q_{\kappa}}$, such that the loaded quality factor $Q$ of the cavity is
\begin{equation}
	Q=\left(\frac{1}{Q_{i}}+\frac{1}{Q_{\kappa}}\right)^{-1}=\frac{Q_{i}}{1+\chi}.
\end{equation}

\subsubsection{Single photon extraction efficiency and indistinguishability}
To determine the single photon extraction efficiency we assume that the emitter is initialized in the excited state with no photons present in the cavity. The problem can then be described in a basis consisting of just 3 states: $\{\ket{1}=\ket{g,0},\ket{2}=\ket{g,1},\ket{3}=\ket{e,0}\}$, respectively corresponding to a state where the emitter is in the ground state and no photons are in the cavity, a state where the emitter is in the ground state and 1 photon is present in the cavity and a state where the emitter is in the excited state and no photon present in the cavity. The rate equations are the same as reported in earlier work. \cite{Sref29,Sref30,Sref31} In the basis $\{\ket{1}=\ket{g,0},\ket{2}=\ket{g,1},\ket{3}=\ket{e,0}\}$ we get
\begin{align}
\frac{d\rho_{22}}{dt}&=-2\Omega\Im(\rho_{23})-\gamma_{p}\rho_{22}\\
\frac{d\rho_{33}}{dt}&=2\Omega\Im(\rho_{23})-\gamma_{e}\rho_{33}\\
\frac{d\Im(\rho_{23})}{dt}&=-\frac{\gamma_{p}+\gamma_{e}+\gamma^{*}}{2}\Im(\rho_{23})+\Omega\rho_{22}-\Omega\rho_{33}
\end{align}
and we assume the system is initially in the excited state, i.e. $\rho(t=0)=\ket{3}\bra{3}$. For a system at low temperature (4K) we can safely assume $\gamma^{*}\ll \gamma_{e}+\gamma_{p}$ such that the $\gamma^{*}$ can be neglected in the equation for $\Im(\rho_{23})$. This is justified as a typical $\gamma^{*}$ at low temperature would be on the order of 10 to 100 GHz \cite{Sref33}, while the total cavity decay rate $\gamma_{p}=\omega_{c}/(2Q)$ would typically be 1000 GHz for a loaded $Q=1000$ (at $\lambda=750$ nm). After solving for $\rho_{22}(t)$, the single photon generation efficiency into the waveguide mode is
\begin{equation}
\eta=\kappa\int_{0}^{\infty}dt\rho_{22}(t)=\frac{\kappa}{(\gamma_{e}+\gamma_{p})\left(1+\frac{\gamma_{e}\gamma_{p}}{4\Omega^{2}}\right)},
\end{equation}
which is the same equation as obtained before. \cite{Sref29,Sref30} If $\gamma^{*}$ can not be neglected, the system can still be solved analytically but the formula becomes quite cumbersome. One could then resort to a full numerical approach as well. The formula for the indistinguishability of photons coupled into the guided mode, derived by Grange et al.\cite{Sref31}, depends on the regime within which the system falls:
\begin{itemize}
\item coherent coupling regime: $2\Omega>\gamma_{p}+\gamma_{e}+\gamma^{*}$:\newline
	\begin{equation}
	V=\frac{(\gamma_{e}+\gamma_{p})(\gamma_{e}+\gamma_{p}+\gamma^{*}/2)}{(\gamma_{e}+\gamma_{p}+\gamma^{*})^{2}}
	\end{equation}
\item incoherent coupling regime: $2\Omega<\gamma_{p}+\gamma_{e}+\gamma^{*}$:
\begin{itemize}
	\item Bad cavity limit: $\gamma_{p}>\gamma_{e}+\gamma^{*}$: \newline
		\begin{equation}
		V=\frac{\gamma_{e}+R}{\gamma_{e}+R+\gamma^{*}}
		\end{equation}
	\item Good cavity limit: $\gamma_{p}<\gamma_{e}+\gamma^{*}$: \newline
		\begin{equation}
		V=\frac{\gamma_{e}+\frac{\gamma_{p}R}{\gamma_{p}+R}}{\gamma_{e}+\gamma_{p}+R}
		\end{equation}
\end{itemize}
\end{itemize}
with
\begin{equation}
R=\frac{4\Omega^{2}}{\gamma_{e}+\gamma_{p}+\gamma^{*}}.
\end{equation}
The above formulas are used to calculate $\eta$ and $\eta V$ as a function of mode volume and $\kappa/\gamma_{c}$ in the main text of the paper.

\subsubsection{Optimum single photon extraction}

Substituting $\kappa=\chi\gamma_{c}$ into the formula for single photon extraction yields
\begin{equation}
\eta=\left(\frac{\chi}{1+\chi+\frac{\gamma_{e}}{\gamma_{c}}}\right)\left(\frac{1}{1+\frac{\gamma_{e}\gamma_{c}}{4\Omega^{2}}(1+\chi)}\right),
\end{equation}
Solving for $d\eta/d\chi=0$ yields the optimum value for $\chi$ to maximize $\eta$ for a given $\Omega^{2}\propto 1/V_{c}$. The optimum reads
\begin{equation}
\chi_{\text{opt}}=\sqrt{\left(1+\frac{\gamma_{e}}{\gamma_{c}}\right)\left(1+\frac{4\Omega^{2}}{\gamma_{e}\gamma_{c}}\right)}=\frac{\kappa_{\text{opt}}}{\gamma_{c}}.
\end{equation}


\end{document}